\documentclass{PoS} \usepackage{wrapfig} \usepackage{amsmath}

\title{{\small{DESY 18-144, DO-TH 18/18}}\\
Strange and non-strange distributions from the collider data}

\ShortTitle{Quark distributions from colliders}

\author{\speaker{Sergey Alekhin}%
         \thanks{This work was supported in part by Bundesministerium f\"ur Bildung und 
                 Forschung (contract 05H15GUCC1),
by the Russian Science Foundation grant No. 14-22-00161 and by Grant No. DE-SC0010073 from the Department of Energy, USA.
}\\ 
II. Institut f\"ur Theoretische Physik, Universit\"at Hamburg,
    Luruper Chaussee 149, D-22761 Hamburg, Germany;\\
        Institute for High Energy Physics,142281 Protvino, Russia\\
        E-mail: \email{sergey.alekhin@desy.de}}

\author{Johannes Bl\"umlein\\ 
        Deutsches Elektronensynchrotron DESY, Platanenallee 6, D--15738 Zeuthen, Germany\\
        E-mail: \email{Johannes.Bluemlein@desy.de}}

\author{Sergey Kulagin\\
        Institute for Nuclear Research of the Russian Academy of Sciences, Moscow 117312, Russia\\
        E-mail: \email{kulagin@ms2.inr.ac.ru}}

\author{Sven-Olaf Moch\\
II. Institut f\"ur Theoretische Physik, Universit\"at Hamburg,
    Luruper Chaussee 149, D-22761 Hamburg, Germany \\      
 E-mail: \email{sven-olaf.moch@desy.de}}

\author{Roberto Petti\\
        Department of Physics and Astronomy, University of South Carolina, Columbia SC 29208, USA\\
        E-mail: \email{roberto.petti@cern.ch}}

\abstract{We check the stability of the ABMP16 fit with respect to 
modifications of quark PDFs suggested in the recent literature: the strange 
sea enhancement and a positive non-vanishing $d/u$ ratio at $x \to 1$. These 
possibilities are examined using test versions of the ABMP16 PDF
fit  which demonstrate no need of those changes. Furthermore, we localize 
peculiar features in other analyses which are responsible for a different behaviour 
of the PDFs obtained.
The strange sea enhancement can be explained by a choice of the 
PDF shapes being not flexible enough and therefore leading to an over-suppressed 
$d$-quark distribution.
This suppression has to be compensated by a corresponding rise by the 
$s$-quark distribution. 
As a result, an unusually large strange sea suppression factor 
is obtained. The non-vanishing 
value of $d/u\vert _{x=1}$ becomes consistent with zero in case the higher-order QCD
corrections to the D{\O} $W$-asymmetry  data, which drive its value, are taken into account. 
Furthermore, the related $e$-asymmetry sample prefers a slightly negative 
value of  $d/u\vert _{x=1}$, although it is consistent with zero. 
These clarifications support confidence in the PDF shapes used in the ABMP16 analyses.}

\FullConference{XXVI International Workshop on Deep-Inelastic Scattering and Related Subjects (DIS2018)\\
		16-20 April 2018\\
		Kobe, Japan}

\begin{document}
\begin{figure}
  \centering
  \includegraphics[width=0.48\textwidth,height=0.45\textwidth]{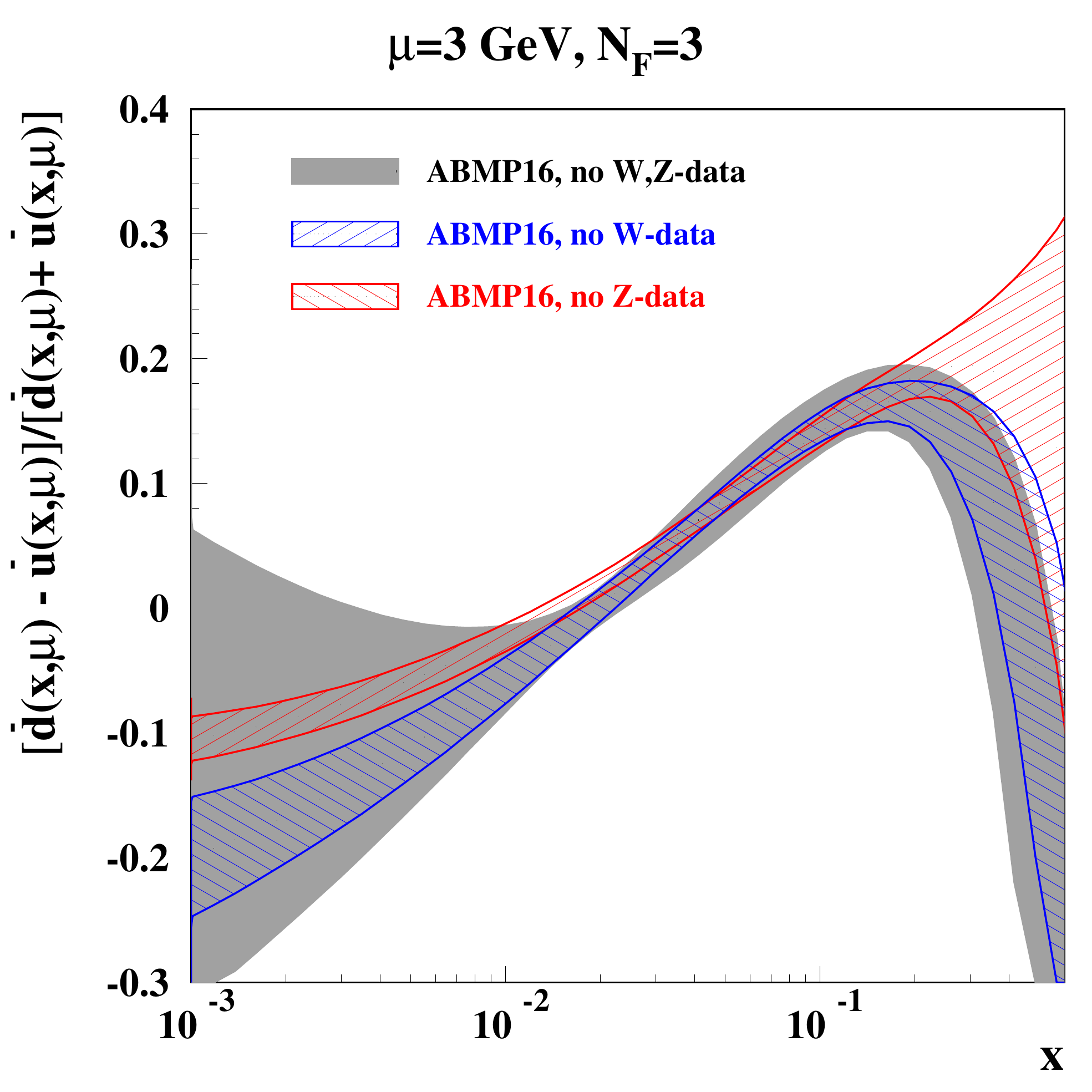}
  \includegraphics[width=0.47\textwidth,height=0.45\textwidth]{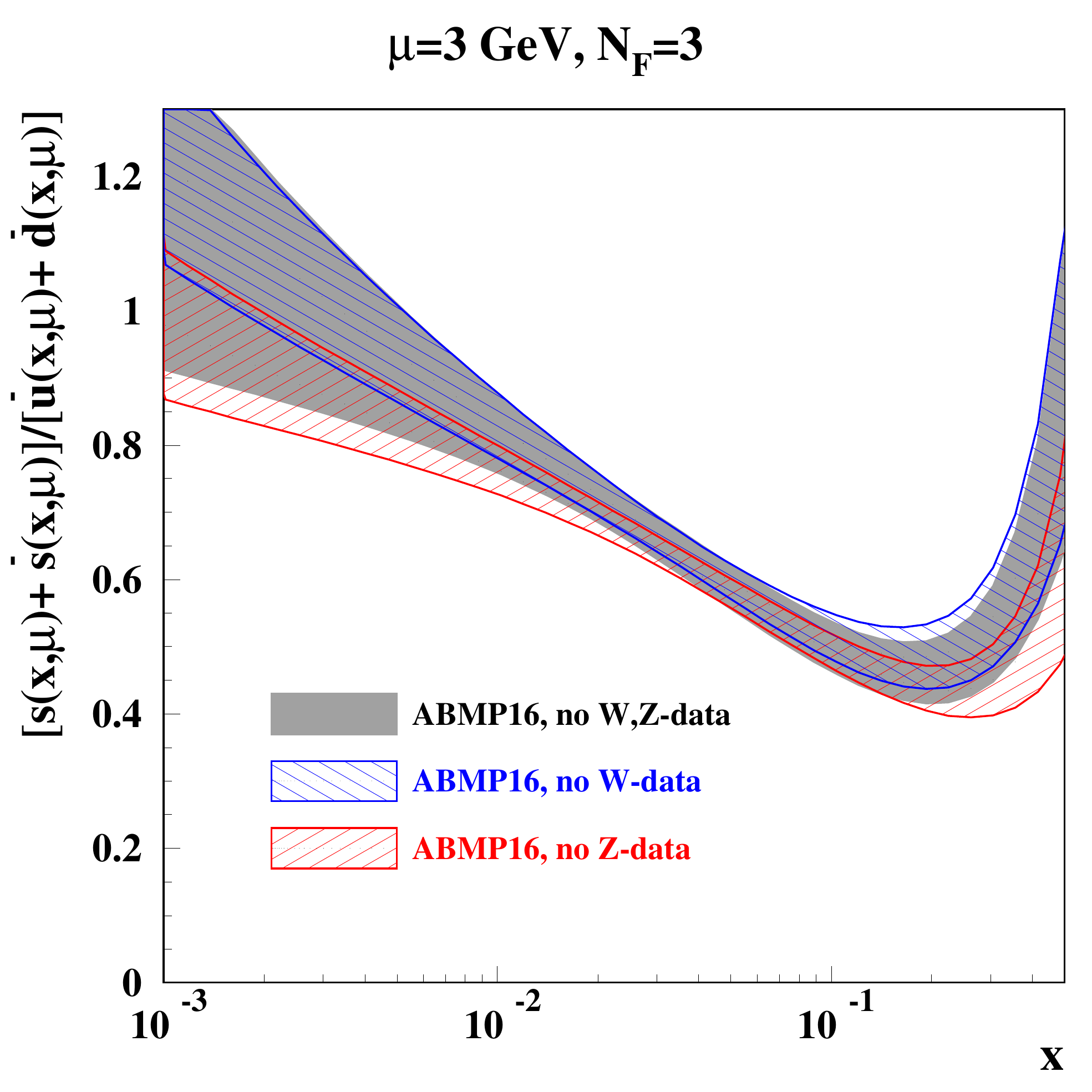}
  \caption{\small
The $1\sigma$ band for the NNLO quark iso-spin asymmetry 
$(\bar{d}-\bar{u})/(\bar{d}+\bar{u})$ (left) and the 
strange sea suppression factor $(s+\bar{s})/(\bar{d}+\bar{u})$ (right)
in the 3-flavor scheme at the scale of $\mu=3~{\rm GeV}$ as a function of 
Bjorken $x$ obtained in the variants of the ABMP16 PDF
fit~\protect\cite{Alekhin:2017kpj} 
with the data on production of $W$-bosons (left-tilted hash), 
$Z$-bosons (right-tilted hash), and both $W$- and $Z$-bosons (shaded area) 
excluded form the fit. 
}
    \label{fig:wz}
\end{figure}

The quark distributions in the nucleon are in general well constrained due to 
the experimental information from deep-inelastic-scattering (DIS)
of leptons off nucleons. Here the most important data are a combination of
inclusive neutral-current (NC) 
scattering data off proton and deuteron targets and data on semi-inclusive 
charged-current (CC) $c$-quark production.  
However, such an approach does not work 
at small Bjorken $x$, where only electron-proton inclusive data, accumulated 
at HERA, are available. On the other hand, the fast emerging samples of the Drell-Yan (DY) 
data, 
collected at the LHC, provide additional constraints on the parton distribution 
functions (PDFs), particularly on the quark ones. These samples,  
together with the DY data obtained at the Tevatron collider in the forward region,
probe PDFs at values of $x$ down to $10^{-4}$ and therefore provide a supplementary 
constraint to the HERA inclusive data. This potential was in particular 
employed in the ABMP PDF fit in order to disentangle small-$x$
$u$- and $d$-quark distributions~\cite{Alekhin:2015cza,Alekhin:2017kpj}. 
The impact of the collider DY data 
on this piece is very significant. Indeed, in the variant of ABMP16 fit, which 
does not include this data sample, the small-$x$ quark distributions are quite
uncertain and the quark iso-spin asymmetry 
$(\bar{d}-\bar{u})/(\bar{d}+\bar{u})$ is in broad 
agreement with 0 at small $x$, while in the nominal ABMP16 fit it is clearly 
negative at $x\sim 10^{-4}$, see~Fig.~\ref{fig:wz}. 
The strange sea determination 
is also improved due to collider DY data, however, in this case the effect 
is less pronounced since the strange sea is constrained to a great extent by 
the neutrino-induced DIS data on charm 
production by the CCFR/NuTeV and NOMAD 
experiments~\cite{Goncharov:2001qe,Samoylov:2013xoa}, which 
are particularly sensitive to the strange sea contribution. 

\begin{figure}
  \centering
  \includegraphics[width=0.48\textwidth,height=0.45\textwidth]{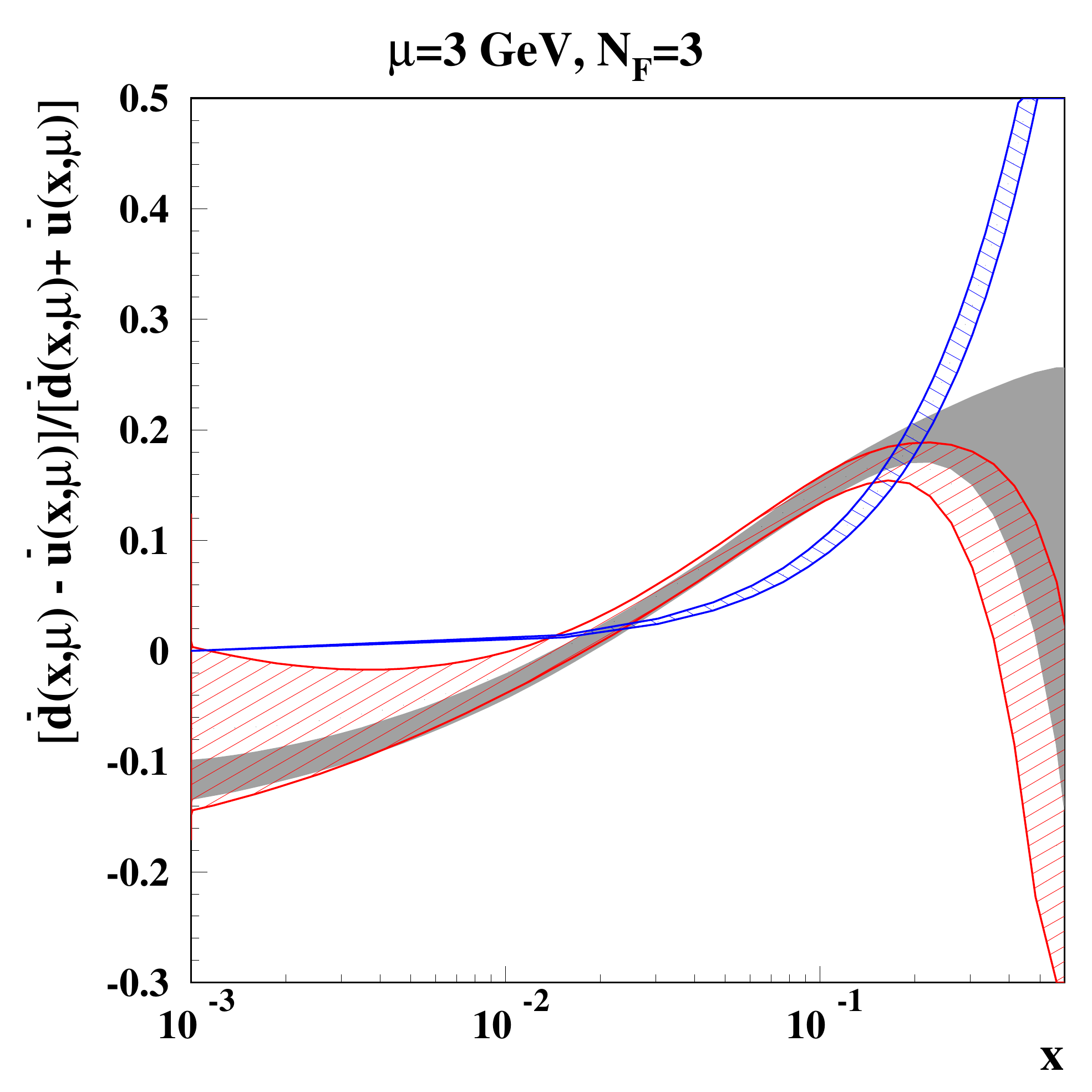}
  \includegraphics[width=0.47\textwidth,height=0.45\textwidth]{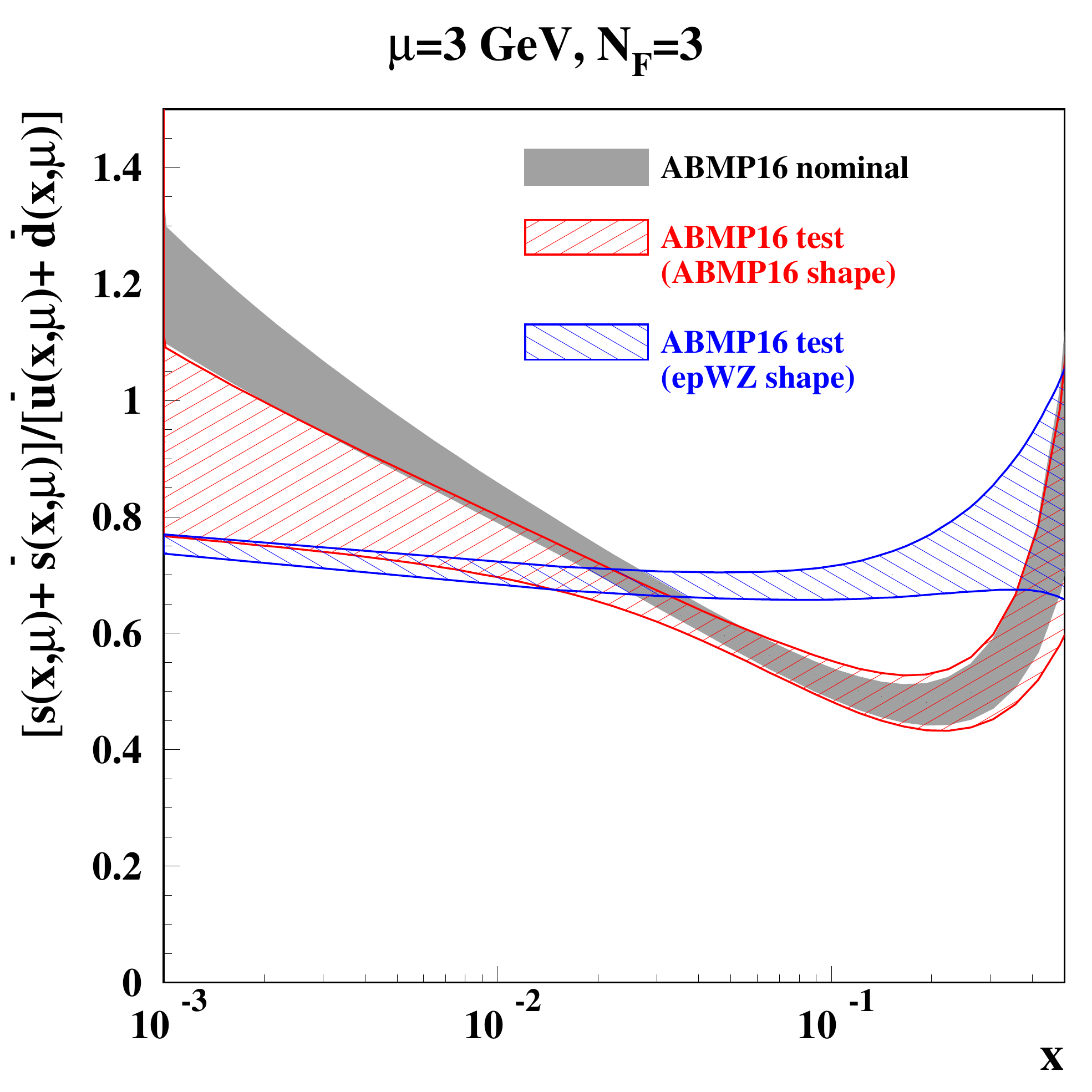}
  \caption{\small
The same as in Fig.\protect\ref{fig:wz} for the test 
variants of the ABMP16 fit with 
the collider DY data excluded, the DIS deuteron data 
added and various PDF shapes used (left-tilted hash: epWZ16 shape, right-tilted 
hash: nominal ABMP16 shape) in comparison with the nominal ABMP16 PDFs
(shaded area).
}
    \label{fig:epwz}
\end{figure}

This is 
in contrast to the PDF fit of Ref.~\cite{Aaboud:2016btc} based on the combination of
the recent high-statistics ATLAS data 
on $W$- and $Z$-boson production with the HERA data on inclusive 
DIS. In the ATLAS analysis 
the strange sea is determined 
by a combination of various constraints coming from the inclusive data and 
as a result it is found to be substantially enhanced as compared to the one 
obtained in the ABMP16 fit. This difference might be in principle explained by  
impact of the ATLAS data~\cite{Aaboud:2016btc},
 which are not included into the ABMP16 
fit, and the different PDF shapes used in these two analyses. The ATLAS fit 
essentially follows the earlier HERAPDF framework derived for the analysis of the HERA 
data alone and, therefore, including many constraints on PDFs, which 
allow a sensible PDF determination using a limited set of the inclusive DIS data. 
To check the impact of such a choice of the PDF shape we perform a variant of the ABMP16 
fit~\cite{Alekhin:2017olj} with the epWZ16 PDF shape, which is employed in the 
ATLAS analysis. Furthermore, 
in order to separate the influence of the experimental input we discard 
all DY collider data used in the nominal ABMP16 analysis, 
including the earlier data from ATLAS \cite{Aad:2011dm}, which are used in 
the ABMP16 analysis, however. At the same time we add to the analysis data on DIS 
off deuteron targets allowing to 
disentangle the quark distributions after discarding the collider DY data.
We account for the deuteron wave-function effects, relevant in this case, by applying 
the Kulagin-Petti model~\cite{Kulagin:2004ie}, which is based on 
the microscopic description in terms of the off-shell nucleon function.
Such an approach provides universal description of the nuclear effects in a wide range of 
targets and the PDFs extracted within this framework are consistent with the 
ones preferred by the existing DY collider data~\cite{Alekhin:2017fpf}. 
It turns out that, despite the ATLAS 
data are not used, the strange sea is still enhanced as compared to 
the nominal ABMP16 PDFs, see Fig.~\ref{fig:epwz}. It is worth noting that 
the charm production data~\cite{Goncharov:2001qe,Samoylov:2013xoa} are 
well described in this variant of fit, with $\chi^2/NDP=167/226$ 
achieved\footnote{The value of $\chi^2$ for the 
CCFR/NuTeV data is smaller than the number of data 
points (NDP) due to the definition of the experimental covariance matrix, which 
includes effects of neighbor bin correlations, so that the
effective number of degrees of freedom is smaller than the number of data points, see 
Ref.~\protect\cite{Goncharov:2001qe} for details.}.
It is interesting to see that the strange sea distribution obtained with the 
ABMP16 PDF shape and the same data selection 
is significantly smaller at $x\sim 0.1$ and agrees with the nominal 
ABMP16 PDFs. However, the value of $\chi^2/NDP=161/226$ obtained 
in this case for the CCFR/NuTeV and NOMAD data is very similar to the
one for the fit with epWZ16 shape. Such a peculiarity is explained 
by the fact that the epWZ16 variant of the fit suggests simultaneous 
suppression of the $d$-quark distribution at $x\sim 0.1$, 
see Fig.~\ref{fig:epwz}. This happens due to the evidently over-constrained fit in 
the epWZ16 shape of the quark iso-spin asymmetry, which by construction 
vanishes at $x\to 0$ and follows this trend at larger $x$. However, this 
is in disagreement with the FNAL-E-866 experiment 
on the muon pair production in proton-proton and proton-deuteron
scattering~\cite{Towell:2001nh}, which clearly reports a positive 
value of $(\bar{d}-\bar{u})$ at this kinematics. As a result, the description of
the E-866 data obtained in the test fit with the epWZ16 shape is quite poor giving 
the value of $\chi^2/NDP=96/39$ to be compared to 49/39 obtained in the 
variant of fit with ABMP16 PDF shape. This brings us to the conclusion that the 
epWZ16 PDF shape is not flexible enough to study the strange sea and tends to 
pull it up at the price of a suppressed  $d$-quark distribution.  

\begin{wrapfigure}{r}{0.5\textwidth}
  \centering
  \includegraphics[width=0.45\textwidth]{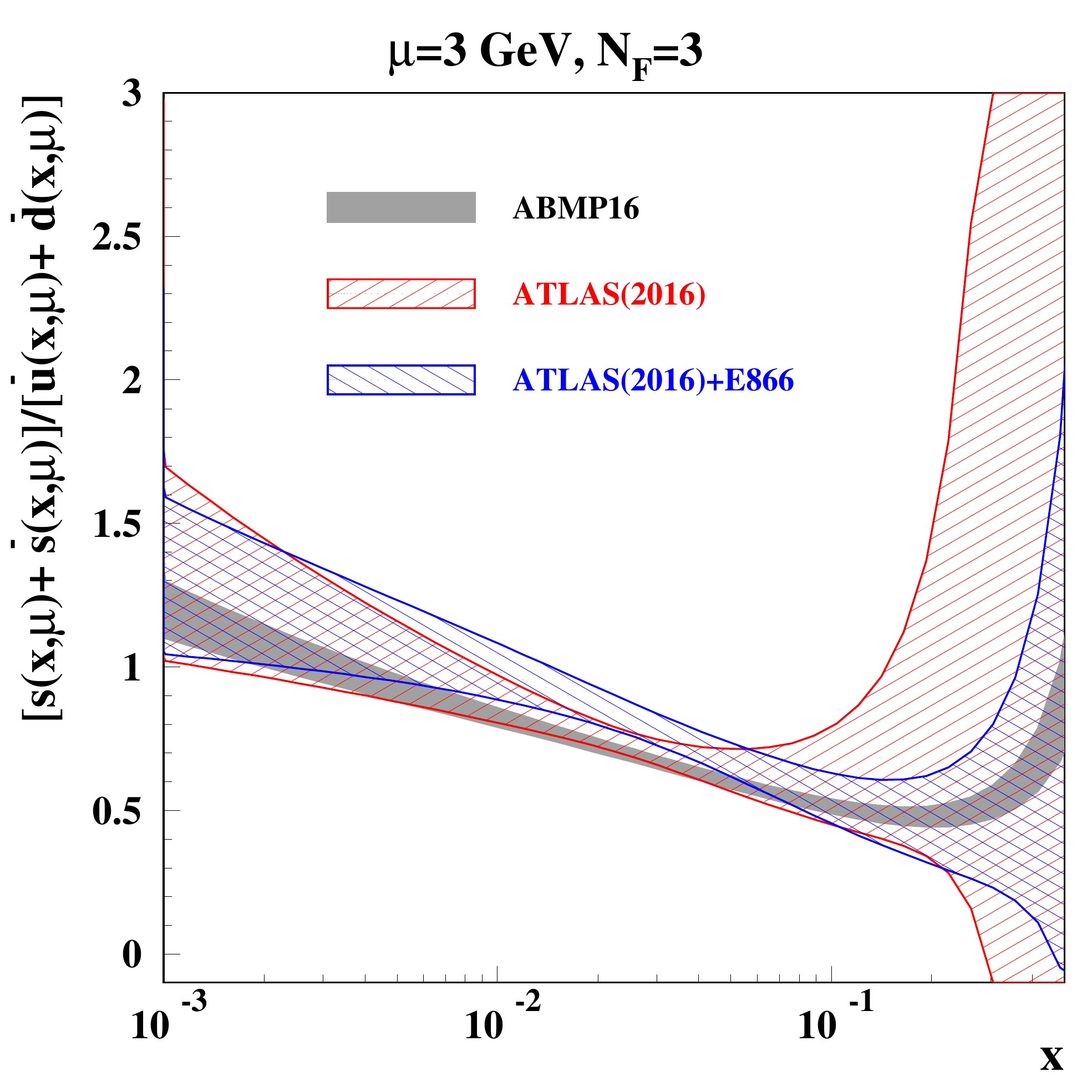}
  \caption{\small The same as in Fig.\protect\ref{fig:wz}
for the strange sea suppression factor $(s+\bar{s})/(\bar{d}+\bar{u})$
obtained in the test variants of the ABMP16 fit with ATLAS data used in 
combination with the inclusive HERA data (left-tilted hash) and the E-866
data on the top (right-tilted hash) in comparison with the nominal ABMP16 PDFs
(shaded area).
}  
  \label{fig:ssupatlas}
\end{wrapfigure}
For a further check of the impact of the ATLAS data on the strange sea
we perform a variant of fit based on  
a combination of the ATLAS and inclusive HERA data
with the ATLAS data on $W$- and $Z$-boson production~\footnote{We take only
the high-statistics $Z$-boson data of the central-rapidity region into account 
and leave out the less accurate data in the forward rapidity region.}.
Such a selection of data 
is in line with the ATLAS analysis of Ref.~\cite{Aaboud:2016btc}. However, 
in this case we use a more flexible ABMP16 PDF shape. The strange sea obtained 
in this way is in a broad agreement with the ATLAS result due to quite large 
uncertainties at $x\gtrsim 0.1$, see Fig.~\ref{fig:ssupatlas}. This means  
the combination of the ATLAS and HERA data is not sufficient to disentangle 
the distributions for
all quark species. Meanwhile, one can improve the fit by adding 
the E-866 data serving as an additional constraint on the quark iso-spin asymmetry  
$(\bar{d}-\bar{u})$. Due to this input the strange sea uncertainty 
reduces substantially, while the central value is still in agreement with 
the ABMP16 result. The latter is driven to a great extent by the 
CCFR/NuTeV and NOMAD data. Here no tension between 
E-866 and ATLAS data is observed, obtaining the values of $\chi^2/NDP=48/39$ and  
40/34, respectively~\footnote{The ATLAS value of $\chi^2$ is  better than the one 
obtained for the respective sample of the ATLAS data in the 
analysis of Ref.~\cite{Aaboud:2016btc}.}. The strange sea distribution obtained in 
this way 
is in reasonable agreement with the one obtained in the ABMP16 fit. The small
enhancement still observed at $x\sim 0.01$ is driven by the $Z$-boson data, 
which deserve further study in view of some tension with the CMS 
results, see Ref.~\cite{Alekhin:2017olj} for details. 

Another debatable issue concerns the ratio $d/u$. Due to the lack of experimental 
data it is 
poorly constrained at $x\sim 1$.
Therefore 
some uncertainty in $\lim_{x\to 1} d/u$ persists. In particular, as it 
was found in a recent NLO analysis by
CJ15~\cite{Accardi:2016qay} $d/u$ is finite at
$x \to 1$, in contrast to the asymptotic value $d/u\to 0$
commonly suggested by various PDF sets available. The CJ15 fit is based on 
a combination of various DIS, DY, and jet production data with the crucial 
role for the determination of $d/u$ at large $x$ played by the D{\O} results on the 
$W$-boson charge asymmetry~\cite{Abazov:2013dsa}. The NLO QCD corrections to 
this process are computed in Ref.~\cite{Accardi:2016qay}
using the $K$-factor approach. However, since for the $\bar{p} p$ initial 
state of the D{\O} experiment the cross sections of $W^+$ and $W^-$ coincide, 
the respective $K$-factors cancel in the 
charge asymmetry and the NLO correction
factors appear equal to one, i.e. just the LO approximation is reproduced in such 
a manner. 
\begin{wrapfigure}{l}{0.5\textwidth}
  \centering
  \includegraphics[width=0.45\textwidth]{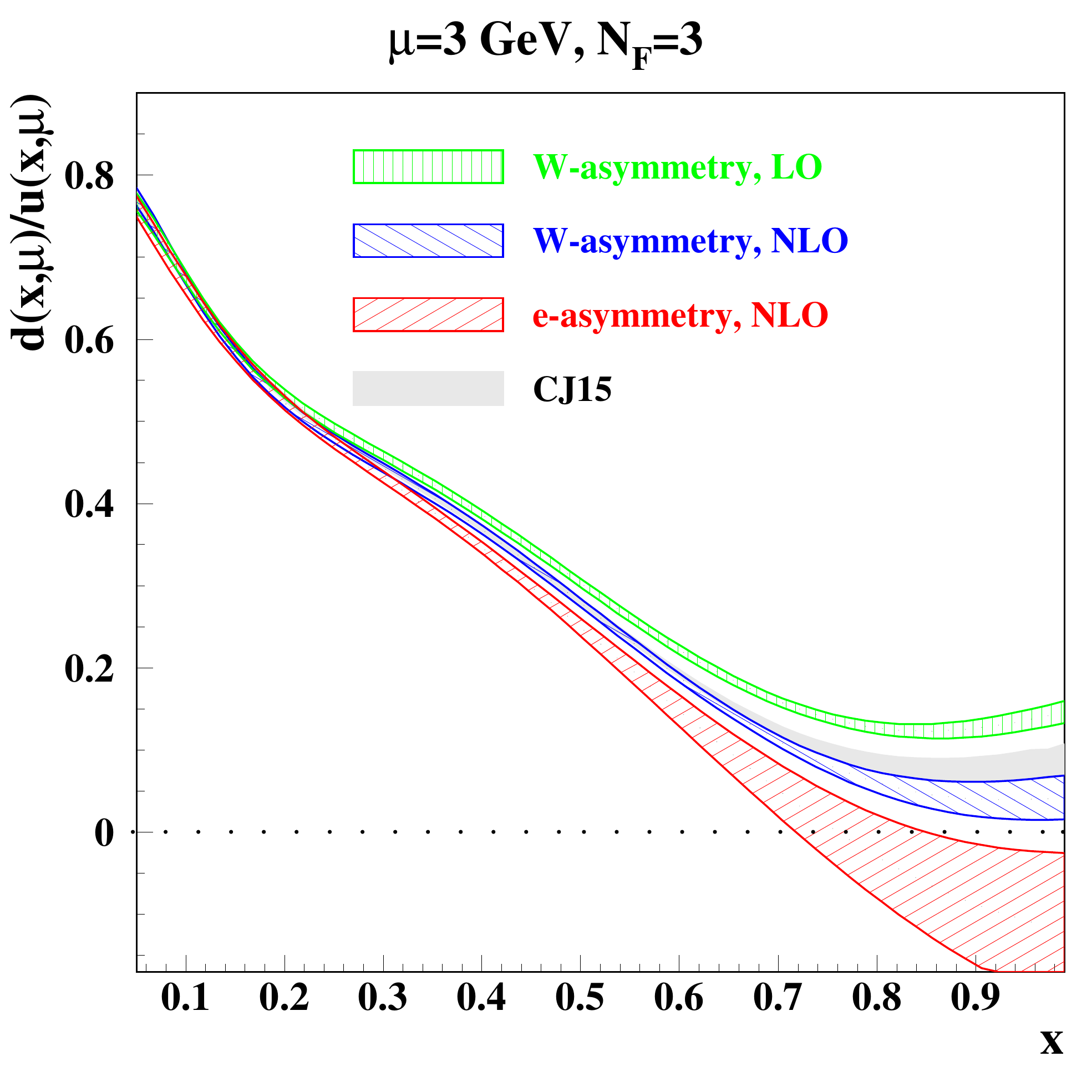}
  \caption{\small The same as in Fig.\protect\ref{fig:epwz}
for the ratio $d/u$ obtained using the CJ15 PDF 
shape~\protect\cite{Accardi:2016qay} and 
with addition of the D{\O} data on 
$W$- and $e$-asymmetry, described within various approximations
(vertical hash: $W$-asymmetry~\protect\cite{Abazov:2013dsa} at LO, 
left-tilted hash: the same in the NLO, 
right-tilted hash: $e$-asymmetry~\protect\cite{D0:2014kma} at NLO)
in comparison with the nominal CJ15 PDFs
(shaded area).
}  
  \label{fig:cj15}
\end{wrapfigure}
To check how this cancellation affects the $d/u$ ratio extracted 
from the $W$-asymmetry data we perform test variants of the fit,  
which reproduce basic features of the CJ15 one. The framework of these tests is
the NLO ABMP16 fit~\cite{Alekhin:2018pai} with all DY collider data dropped
and the D{\O} $W$-asymmetry data included instead. Such a selection allows to 
illustrate the impact of this data set on the $d/u$ ratio and 
to investigate the significance of its treatment. 

We also add the deuteron fixed-target data in order to allow to
disentangle the quark distributions at moderate values of $x$. Moreover, in 
this way we 
provide better consistency with the CJ15 data selection. The PDFs are 
parameterized as in the CJ15 fit, however, with two unessential  
modifications. Firstly, we release the strange sea suppression factor
providing in this way a better description of the CCFR/NuTeV and NOMAD data. 
Secondly, we reduce the number of parameters responsible for the 
large-$x$ gluon behavior since the data sets used are not sensitive to this 
region. 

The ratio $d/u$ obtained in the variant of this test fit with the 
LO approximation employed for the description of the $W$-asymmetry data 
demonstrates a rising trend at large $x$ resulting into 
$(d/u)\vert _{x=1}=0.15\pm0.01$, 
 in line with the CJ15 finding, see Fig.~\ref{fig:cj15}.
However, with account of the NLO corrections it takes essentially lower values 
with $(d/u)\vert _{x=1}=0.04\pm0.03$. This is almost consistent with zero within 
$1\sigma$. Note that the uncertainties in $d/u$ rise essentially from LO 
to NLO. This is obviously related to the involvement of gluon-initiated 
processes appearing in the latter case and the impact of the gluon distribution 
uncertainty propagating through this channel. One more aspect of this study
is a selection between the  D{\O} data on the $W$-asymmetry and 
the $e$-asymmetry~\cite{D0:2014kma}. 
Both sets are derived from the same sample of events and the latter is, in fact,
directly measured in experiment, while the $W$-asymmetry is 
obtained by processing this primary information. Such a processing requires 
a certain modeling and the result of this is sensitive   
to the PDFs used. Moreover, the PDFs obtained from the analysis of the 
$e$-asymmetry data are not consistent with the data on the $W$-asymmetry, see
Fig. 9 in Ref.~\cite{Alekhin:2015cza}. Motivated by this bias we also
perform a variant of test fit with the $e$-asymmetry data instead 
of the $W$-asymmetry. The NLO corrections are applied in order to provide 
the relevant theoretical accuracy. The $d/u$ ratio obtained in this way goes 
further down with even larger uncertainties, see Fig.~\ref{fig:cj15},    
and the value $(d/u)\vert _{x=1}=-0.12\pm0.09$ is consistent with zero within 
the uncertainties. Comparing it with the value obtained from the $W$-asymmetry
we find a broad agreement, however, the $e$-asymmetry result is less sensitive 
to model assumptions, as we mentioned before. In any case, the CJ15 result on 
$d/u$  at large $x$ is obviously due to a particular selection of the 
data and the limited theoretical accuracy of the analysis. Otherwise we arrive
at the  value of $(d/u)\vert _{x=1}$ consistent with zero which justifies such a 
constraint commonly imposed in the PDF fits.

In conclusion, we considered modifications of quark PDFs suggested in the 
recent literature: the strange sea enhancement~\cite{Aaboud:2016btc} and 
a positive non-vanishing $d/u$ ratio at $x\to 1$~\cite{Accardi:2016qay}. These 
possibilities were examined using test versions of the ABMP16 PDF
fit~\cite{Alekhin:2017kpj}, which demonstrate no need of those changes.
Furthermore, we localize peculiar features of these analyses 
which are responsible for the uncommon behavior of their PDFs obtained.
The strange sea enhancement~\cite{Aaboud:2016btc}
 can be explained by non-flexible 
PDF shapes leading to an over-suppressed $d$-quark distribution.
This suppression has to be compensated by the respective rise of the 
$s$-quark distribution. 
As a result an unusually  large strange sea suppression factor 
is obtained. The non-vanishing 
value of $d/u\vert _{x=1}$~\cite{Accardi:2016qay}
becomes consistent with zero in case the NLO corrections to the  D{\O} $W$-asymmetry  
data are taken into account. Furthermore, 
the related $e$-asymmetry sample prefers a slightly negative 
value of  $d/u\vert _{x=1}$, although being consistent with zero.
These clarifications support confidence in the PDF shape used in the ABMP16 fit.


\begin{thebibliography}{99}
\bibitem{Alekhin:2015cza}
  S.~Alekhin, J.~Bl\"{u}mlein, S.~Moch and R.~Pla\v cakyt\. e,
  Phys.\ Rev.\ D {\bf 94} (2016)  114038
  [arXiv:1508.07923 [hep-ph]].
  %%CITATION = doi:10.1103/PhysRevD.94.114038;%%
  %28 citations counted in INSPIRE as of 06 Aug 2018

\bibitem{Alekhin:2017kpj}
  S.~Alekhin, J.~Bl\"{u}mlein, S.~Moch and R.~Pla\v cakyt\. e,
  Phys.\ Rev.\ D {\bf 96} (2017)  014011 
 [arXiv:1701.05838 [hep-ph]].
  %%CITATION = doi:10.1103/PhysRevD.96.014011;%%
  %87 citations counted in INSPIRE as of 06 Aug 2018

\bibitem{Goncharov:2001qe}
  M.~Goncharov {\it et al.} [NuTeV Collaboration],
  Phys.\ Rev.\ D {\bf 64} (2001) 112006
  [hep-ex/0102049].
  %%CITATION = doi:10.1103/PhysRevD.64.112006;%%
  %285 citations counted in INSPIRE as of 06 Aug 2018

\bibitem{Samoylov:2013xoa}
  O.~Samoylov {\it et al.} [NOMAD Collaboration],
  Nucl.\ Phys.\ B {\bf 876} (2013) 339
  [arXiv:1308.4750 [hep-ex]].
  %%CITATION = doi:10.1016/j.nuclphysb.2013.08.021;%%
  %37 citations counted in INSPIRE as of 06 Aug 2018

\bibitem{Aaboud:2016btc}
  M.~Aaboud {\it et al.} [ATLAS Collaboration],
  Eur.\ Phys.\ J.\ C {\bf 77} (2017)  367 [arXiv:1612.03016 [hep-ex]].
  %%CITATION = doi:10.1140/epjc/s10052-017-4911-9;%%
  %58 citations counted in INSPIRE as of 06 Aug 2018

\bibitem{Aad:2011dm}
  G.~Aad {\it et al.} [ATLAS Collaboration],
  Phys.\ Rev.\ D {\bf 85} (2012) 072004
  [arXiv:1109.5141 [hep-ex]].
  %%CITATION = doi:10.1103/PhysRevD.85.072004;%%
  %342 citations counted in INSPIRE as of 06 Aug 2018

\bibitem{Alekhin:2017olj}
  S.~Alekhin, J.~Bl\"{u}mlein and S.~Moch,
  Phys.\ Lett.\ B {\bf 777} (2018) 134
  [arXiv:1708.01067 [hep-ph]].
  %%CITATION = doi:10.1016/j.physletb.2017.12.024;%%
  %7 citations counted in INSPIRE as of 08 Aug 2018

\bibitem{Kulagin:2004ie}
  S.~A.~Kulagin and R.~Petti,
  Nucl.\ Phys.\ A {\bf 765} (2006) 126
  [hep-ph/0412425].
  %%CITATION = doi:10.1016/j.nuclphysa.2005.10.011;%%
  %194 citations counted in INSPIRE as of 16 Aug 2018

\bibitem{Alekhin:2017fpf}
  S.~I.~Alekhin, S.~A.~Kulagin and R.~Petti,
  Phys.\ Rev.\ D {\bf 96} (2017)  054005
  [arXiv:1704.00204 [nucl-th]].
  %%CITATION = doi:10.1103/PhysRevD.96.054005;%%
  %5 citations counted in INSPIRE as of 07 Aug 2018

\bibitem{Towell:2001nh}
  R.~S.~Towell {\it et al.} [NuSea Collaboration],
  Phys.\ Rev.\ D {\bf 64} (2001) 052002
  [hep-ex/0103030].
  %%CITATION = doi:10.1103/PhysRevD.64.052002;%%
  %399 citations counted in INSPIRE as of 09 Aug 2018

\bibitem{Accardi:2016qay}
  A.~Accardi, L.~T.~Brady, W.~Melnitchouk, J.~F.~Owens and N.~Sato,
  Phys.\ Rev.\ D {\bf 93} (2016)  114017
  [arXiv:1602.03154 [hep-ph]].
  %%CITATION = doi:10.1103/PhysRevD.93.114017;%%
  %89 citations counted in INSPIRE as of 14 Aug 2018

\bibitem{Abazov:2013dsa}
  V.~M.~Abazov {\it et al.} [D0 Collaboration],
  Phys.\ Rev.\ Lett.\  {\bf 112} (2014) no.15,  151803
   Erratum: [Phys.\ Rev.\ Lett.\  {\bf 114} (2015)  049901]
  [arXiv:1312.2895 [hep-ex]].
  %%CITATION = doi:10.1103/PhysRevLett.114.049901, 10.1103/PhysRevLett.112.151803;%%
  %22 citations counted in INSPIRE as of 14 Aug 2018

\bibitem{D0:2014kma}
  V.~M.~Abazov {\it et al.} [D0 Collaboration],
  Phys.\ Rev.\ D {\bf 91} (2015) no.3,  032007
   Erratum: [Phys.\ Rev.\ D {\bf 91} (2015)  079901]
  [arXiv:1412.2862 [hep-ex]].
  %%CITATION = doi:10.1103/PhysRevD.91.032007, 10.1103/PhysRevD.91.079901;%%
  %35 citations counted in INSPIRE as of 14 Aug 2018

\bibitem{Alekhin:2018pai}
  S.~Alekhin, J.~Bl\"{u}mlein and S.~Moch,
  Eur.\ Phys.\ J.\ C {\bf 78} (2018)  477
  [arXiv:1803.07537 [hep-ph]].
  %%CITATION = doi:10.1140/epjc/s10052-018-5947-1;%%
  %5 citations counted in INSPIRE as of 14 Aug 2018

\end{thebibliography}
\end{document}